\documentclass{article}

\usepackage[dvips]{graphicx}
\usepackage{float}
\usepackage{url}
\usepackage{cite}
\usepackage{authblk}
\title{\textbf{Exotic Higgs Decay $h \rightarrow a_1 a_1$ at the International Linear Collider: a Snowmass White Paper}}

\author[1]{Tao Liu}
\author[2]{C.T. Potter}
\affil[1]{The Hong Kong University of Science and Technology}
\affil[2]{Department of Physics, University of Oregon}

\date{\today}

\begin{document}

\maketitle

\begin{abstract}
A Higgs factory like the International Linear Collider (ILC) can play a significant role in searching for exotic decays of Higgs bosons. As an illustration, we investigate the ILC sensitivity for the decay topology $h\to a_1 a_1 \to \tau\bar  \tau\tau \bar \tau$ in the Next-to-Minimal-Supersymmetric-Standard-Model (NMSSM). Here $h$ can be either Standard-Model-like or non-standard, and $a_{1}$ is the lightest CP-odd Higgs boson. We also compare results to expectations for this channel at the LHC. 
\end{abstract}

\section{Introduction \label{section1}}
The discovery of a Higgs-like resonance at the CMS~\cite{Chatrchyan:2012ufa} and ATLAS~\cite{Aad:2012tfa} heralds the beginning of a new era of Higgs physics. The Higgs in the Standard Model (SM) suffers divergent quantum corrections to its mass, caused by the large hierarchy between the electroweak (EW) scale and the Planck scale. In most new physics scenarios addressing the gauge hierarchy problem, the Higgs mass stabilization mechanism manifests itself through Higgs couplings absent in the SM. In addition, the Higgs is one of the two SM fields that can have renormalizable couplings to SM singlet operators~\cite{Patt:2006fw}. The Higgs therefore may be the main window into new physics and systematic studies on the Higgs properties should be pursued, e.g, coupling extractions and exotic decay searches. 

Exotic Higgs decays are particularly interesting because any signal would be an unambiguous signature for physics beyond the SM (BSM). Due to the small SM Higgs decay width ($\Gamma \sim 4$ MeV for $m_h \sim 125$ GeV), a small coupling between the Higgs boson and some light particles may yield a large exotic Higgs decay branching fraction. Currently, ATLAS reports an upper limit $Br(h_{125} \rightarrow invisible) < 65$\% at 95\% C.L. (expected $Br(h_{125} \rightarrow invisible) < 85$\% at 95\% C.L.) \cite{ATLAS-CONF-2013-011}. Similarly CMS reports upper limit observed 75\% at 95\% C.L. (expected 91\% at 95\% C.L.) \cite{CMS-PAS-HIG-13-018}. Because the LHC lacks sensitivity in measuring the Higgs-glue-glue coupling directly, it is very difficult to constrain the upper bound for Br$(h\to exotic)$ below $10\%$, even with the full 300 fb$^{-1}$ data of the LHC14~\cite{Peskin:2012we}. Therefore, exotic Higgs decays provide  a very effective tool to explore potential new physics couplings to the Higgs boson.

The LHC is expected to be upgraded to its designed beam energy $13-14$~TeV at the end of 2014, and to collect up to 300~fb$^{-1}$ during Run 2. This provides an opportunity for searching for exotic Higgs decays. Motivated by this, LHC studies have been or are being pursued by theorists in various contexts.  However, though the LHC may play a significant role in exploring some exotic Higgs decays, its sensitivity is weak in some cases. In one case the  SM-like Higgs decays into soft or collimated $\tau$ leptons or $b$ quarks, with or without missing particles. In a second case the Higgs decays to purely missing particles. To have sensitivity to such searches, it is typical that $\sim \mathcal O(100-1000)$~fb$^{-1}$ LHC14 data  is required, given Br$(h\to {\rm \emph{exotic}}) > 10\%$. This is either because of the hadronic collider environment at the LHC or due to the lack of kinematic handles in the final states. If the branching ratios of such exotic decays are below $10\%$, discovery may be beyond the reach of the LHC14.

In these cases a Higgs factory like the International Linear Collider (ILC) is invaluable. One motivation for constructing such a machine is that it can precisely  measure the Higgs couplings, including the Higgs-glue-glue and Higgs self couplings. Solving the hierarchy problem typically requires BSM physics to enter the effective theory at TeV scale. If so, the Higgs couplings in the SM are expected to have a  deviation of $\mathcal O(1\%)$ level~\cite{Peskin:2012we}. Unless significant improvement can be achieved for suppressing systematic uncertainties, it is very difficult for the LHC14 to reach such sensitivity. The ILC however can do much better. It is expected to be able to measure a deviation of $\mathcal O(1\%)$ level, and hence to probe TeV scale BSM physics, assuming reasonable integrated luminosity~\cite{Peskin:2012we}. 

In this work, we show that exotic Higgs decays provide another case, justifying the value of a Higgs machine like the ILC. The ILC is a machine not only for precise measurements, but also for discoveries. Though exotic Higgs decays can occur in many contexts, for the consideration of representativity, we will work in the Next-to-minimal-Supersymmetric-SM (NMSSM). The R-~\cite{Dobrescu:2000jt,Dobrescu:2000yn,Dermisek:2005ar} and PQ-symmetry limits~\cite{Draper:2010ew}, in the NMSSM provides supersymmetric benchmark for various exotic Higgs decays. As an illustration, we consider a specific case in the R-symmetry limit of the NMSSM where $h_{1,2} \to 2 a_1$ are significant~\cite{Belanger:2012tt} and $a_1$ is the lightest CP-odd Higgs boson and serves as an R-axion, $a_1$ dominantly decays into a pair of $\tau$ leptons.  Since neither CMS nor ATLAS report searches in the channel $h_{125} \rightarrow 2 a_1$, these decays may proceed with high branching ratio yet still go undetected. Then we compare between the LHC and the ILC performance. We will show that the ILC can serve as a discovery machine for exotic Higgs decays which are challenging for the LHC14.

\section{NMSSM Higgs Parameters \label{sec:nmssm}}

A review of the Higgs sector of the NMSSM can be found in \cite{Ellwanger:2009dp}. Briefly, its superpotential and softly SUSY breaking terms are given by 
\begin{eqnarray}
\mathbf{W}&=&\lambda \mathbf{S}\mathbf{H_u}\mathbf{H_d} + \frac{\kappa}{3} \mathbf{S}^3, \nonumber \\
 V_{\it soft} &= &{m^2_{H_d}} |H_d|^2 + {m^2_{H_u}} |H_u|^2 +{m^2_S}|S|^2  + (- \lambda A_{\lambda} H_u H_d S + \frac{1}{3} A_\kappa \kappa S^3 + h.c.).
\end{eqnarray}
Here $H_d$, $H_u$ and $S$ denote the neutral scalar fields in the ${\bf  H_d}$, ${\bf H_u}$ and ${\bf S}$ supermultiplets, respectively.  Once the singlet scalar $S$ obtain a VEV $\langle S \rangle$, an effective $\mu$ parameter $\mu_{\rm eff} = \lambda \langle S \rangle$ can
be generated. The NMSSM Higgs sector is determined by six free parameters at tree level: $\lambda, \kappa, A_{\lambda}, A_{\kappa}, \tan \beta$ and $\mu_{eff}$. In addition to the Higgs spectrum of the MSSM, the NMSSM contains one extra CP-even $h$ and one extra CP-odd scalar $a$. With subscripts denoting mass ordering, the NMSSM Higgs sector includes neutral CP-odd $a_1, a_2$, neutral CP-even $h_1, h_2, h_3$ and charged $H^+, H^-$.

To illustrate the sensitivity of the ILC, we consider the exotic decay mode of the SM-like Higgs in R-symmetry limit of the NMSSM
\begin{eqnarray}
h \to a_1 a_1 \to \tau \bar \tau  \tau \bar \tau \ .
\label{decay}
\end{eqnarray}
with $2 m_{a_1}<m_h$. In addition, it is pointed out recently that~\cite{Belanger:2012tt} this decay topology can be applied to 
explain the LEPII $2 \sigma$ excess near $m_{b \bar{b}} \approx 90-100$~GeV in the $Zb\bar{b}$ channel. In this scenario $h_{125}$, the 125~GeV boson recently observed at the LHC is identified with the NMSSM $h_2$. While signal strengths in various decay channels reported from CMS and ATLAS are consistent with the SM signal strengths, they are also  consistent with a large branching ratio to invisible (or undetected) final states. In this scenario, the $h_1$ is responsible for the LEPII excess,  with $h_1 \rightarrow b\bar{b}$ suppressed to some extent by turning the exotic decay mode in Eq.(\ref{decay})~\cite{Belanger:2012tt}. For $m_{a_1}>2m_{B}$, $a_1 \rightarrow b \bar{b}$ dominates. Limits on $h_1 \rightarrow 2 a_1\rightarrow b\bar{b}b\bar{b}$ from LEPII rule out $m_{h_1}<110$~GeV for $m_{a_1}>2m_{B}$ \cite{Schael:2006cr}. So $2 m_{\tau} < m_{a_1} < 2m_{B}$ and $m_{h_1} \approx 90-100$~GeV are suggested for explaining the LEPII excess.  The most constraining limits on this scenario are set by the ALEPH collaboration \cite{Schael:2010aw}. While neither ATLAS nor CMS has reported searches for this scenario, the LHC sensitivity is studied in \cite{Lisanti:2009uy, Belyaev:2008gj, Englert:2011iz, Cerdeno:2013cz}.

We seek NMSSM Higgs model parameters $\lambda, \kappa, A_{\lambda}, A_{\kappa}, \tan \beta$ and $\mu_{eff}$ which yield $m_{a_1} \approx 10$~GeV, $m_{h_1} \approx 90-100$~GeV and $m_{h_2} \approx 125$~GeV. Radiative corrections in the Higgs sector require a full specification in other sectors. We use NMSSMTools 3.2.4 \cite{Ellwanger:2004xm,Ellwanger:2005dv,Belanger:2005kh} to calculate the mass spectrum, widths and branching ratios. See Table \ref{tab:params} for the chosen parameters and resulting masses and branching ratios. The value of $A_{\lambda}$ is determined by the parameter $m_{A}= \frac{\lambda v_s}{\sin 2 \beta} ( \sqrt{2} A_{\lambda} + \kappa v_s) $ where $v_s = \sqrt{2} \langle S \rangle$. This model contains all of the interesting phenomenology described in Section \ref{section1}, namely $h_1 \rightarrow 2 a_1$ and $a_1 \rightarrow \tau^+ \tau^-$ dominant with $m_{a_1} \approx 10$~GeV, $m_{h_1} \approx 90-100$~GeV and $m_{h_2} \approx 125$~GeV. NMSSMTools reports that the model predicts values for $b \rightarrow s \gamma$, $B_s \rightarrow \mu^+ \mu^-$, $B \rightarrow \tau \nu$ and the anomalous magnetic moment $\Delta a_{\mu}$ within experimental constraints. Moreover the lightest chargino mass $m_{\chi_1^+}$ is just at the PDG limit and the dark matter relic density is near the lower allowed limit. From this parameter benchmark  we can easily rescale the ILC sensitivities to other parameter regions, using the results obtained in this analysis.

\begin{table}[tbp]
\begin{center}
\begin{tabular}{||c|c||c|c||c|c||} \hline
Parameter & Value & Scalar & Mass [GeV] & Decay & Br [\%] \\ \hline
$\lambda$ & 0.3 & $a_1$ & 10.3 & $h_1 \rightarrow 2a_1$ & 85.4 \\
$ \kappa$ & 0.1 & $h_1$ & 91.6 & $h_2 \rightarrow 2a_1$ & 87.4 \\
$A_{\kappa}$ & 11.6 & $h_2$ & 124.5 & $a_1 \rightarrow \tau^+\tau^-$ & 73.2 \\
$m_{A}$ & 465 GeV & $a_2$ & 465.2 & $a_1 \rightarrow 2g$ & 22.3 \\
$\tan \beta$ & 3.1 & $h_3$ &  469.2 & $a_1 \rightarrow c\bar{c}$ & 3.1 \\
$\mu_{eff}$ & 165 GeV & $H^{\pm}$ & 465.7 & $a_1 \rightarrow \mu^+ \mu^-$ & 0.3 \\ \hline
\end{tabular}
\caption{NMSSM parameters with Higgs  mass spectrum and dominant branching ratios.}   
\label{tab:params}
\end{center}
\end{table}

\section{ILC Analysis and Results}

\subsection{SiD Detector and Event Simulation}

The SiD detector comprises a compact vertex detector instrumented with silicon pixels for vertex reconstruction, a main tracker instrumented with silicon strips for measuring charged particle momentum, an electromagnetic calorimeter with silicon  strips in the active layers and Tungsten in the passive layers for measuring electromagnetic energy deposits, a hadronic calorimeter with glass resistive plate chambers in the active layers and steel in the passive layers for measuring hadronic energy deposits, and a muon system instrumented with scintillators in the iron flux return of a 5T solenoidal magnet. Full details of the SiD design can be found in Volume 4 of the ILC Technical Design Report \cite{Adolphsen:1541315prime}. 

Generation of the signal process $e^+ e^- \rightarrow Zh_{1,2} \rightarrow f\bar{f}2 a_1$ was performed with the Whizard event generator \cite{Kilian:2007gr,Moretti:2001zz}, which has a full implementation of the NMSSM \cite{Reuter:2009ex}. Whizard interfaces the NMSSM model described in Section \ref{sec:nmssm} with the SLHA \cite{Allanach:2008qq} file generated by NMSSMTools. Signal events are weighted by production cross section multiplied by the branching ratio for $Z \rightarrow f \bar{f}$. For a pure polarization state $e_{L}^{-}e_{R}^{+}$ ($e_{R}^{-}e_{L}^{+}$), the $\sqrt{s}=250$~GeV cross section reported by Whizard for $Zh_1 \rightarrow \mu^+ \mu^- 2 a_1$ is $17.034 \pm 0.006$ fb ($13.105 \pm 0.005$ fb). For $Zh_2 \rightarrow \mu^+ \mu^- a_1 a_1$ it is $5.715 \pm 0.002$ fb ($4.397 \pm 0.002$ fb).
After weighting the event yields correspond to integrated luminosities of 250fb$^{-1}$ for $\sqrt{s}=250$~GeV. Generation of all SM backgrounds is also performed with Whizard. The dominant background to $h_{1,2} \rightarrow 2 a_1\rightarrow 4\tau_{1-pr}$ is the process $e^+ e^- \rightarrow ZZ \rightarrow \mu^+ \mu^- \tau_{1-pr} \tau_{3-pr}$, so a high-statistics sample of this background is produced with Whizard.

Details of the full SiD detector simulation and event reconstruction can be found in \cite{Adolphsen:1541315prime}. After event generation, signal and background events are passed through a detector simulation with SLIC, a program with full GEANT4 \cite{Agostinelli2003250prime} functionality. Energy deposits expected from generator particles are simulated in sensitive regions of the detector subsystems and are then digitized. Particles are reconstructed as particle flow objects using particle flow algorithms, which improves jet resolution.

 \begin{figure}[t]
  \centering
   \includegraphics[width=0.495\textwidth, draft=false]{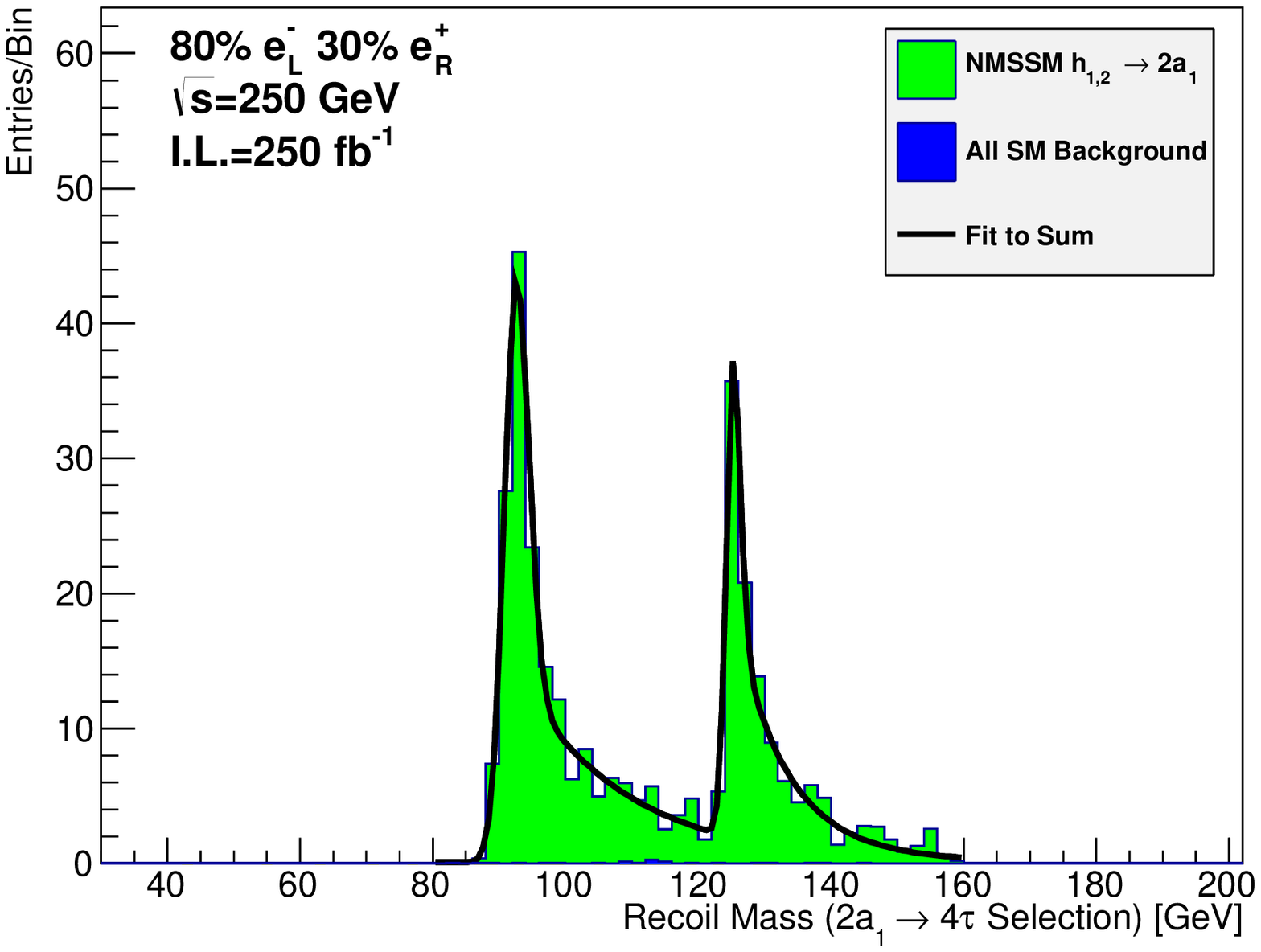}
   \includegraphics[width=0.495\textwidth, draft=false]{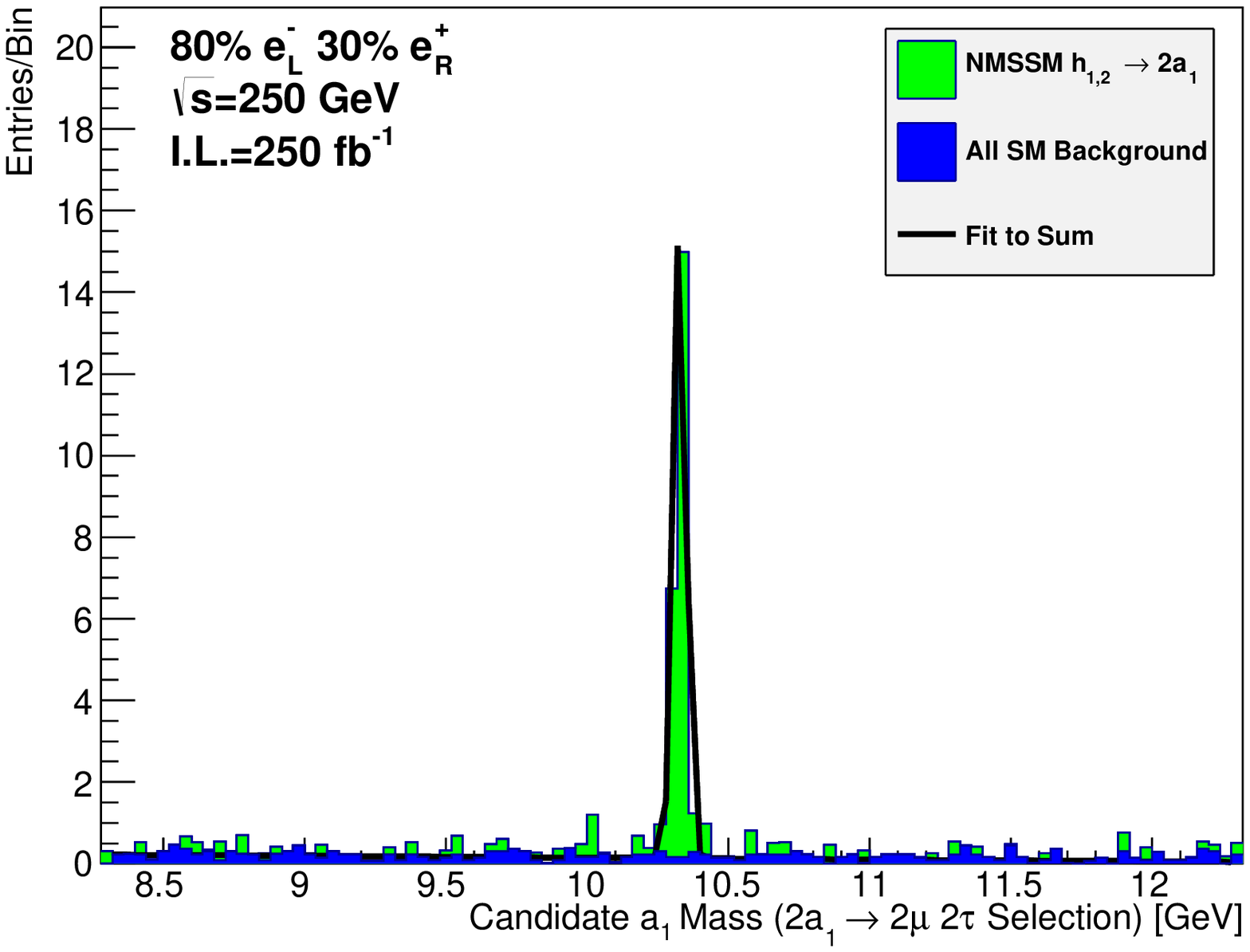}
  \caption{At left, the fit to the $h_{1,2}$ recoil mass distribution after full $4\tau$ analysis selection. At right, the fit to the reconstructed $a_1 \rightarrow \mu^+ \mu^-$ mass distribution after the full $2\mu 2\tau$ selection. The plots assume $\sqrt{s}=250$~GeV, $250$fb$^{-1}$ integrated luminosity, and 80\% $e_{L}^{-}$, 30\% $e_{R}^{+}$ beam polarization.}
  \label{fig:plots}
\end{figure}

\subsection{The $h_{1,2} \rightarrow 2a_1 \rightarrow 4\tau$ Channel}

The data analysis selection seeks to identify the dominant decays of the $h_{1,2}$ in the recoil of $Z \rightarrow \mu^+ \mu^-$. Since the decay of $\tau$ to one-prongs ($e, \mu, \pi$) is dominant we identify $h_{1,2} \rightarrow 2 a_1\rightarrow 4\tau$ as four-track events in the recoil of the $Z$ with net charge zero. The selection requirements are as follows:

\begin{itemize}

\item require at least two muons with $p_T>5$~GeV ($N_{\mu5} \geq 2$)

\item require the muon pair closest to the $Z$ mass within $3\sigma$ of the nominal $Z$ mass ($\vert m_{Z}-m_{\mu^+ \mu^-} \vert < 3\sigma$)

\item require exactly six tracks with $p_T>0.2$~GeV ($N_{trk}=6$)

\item require zero net charge in the recoil tracks ($Q_{4trk}=0$)

\item veto $\tau \rightarrow a_1(1260) \nu$ by requiring candidate $a_1(1260)$ mass $m_{3trk}>2$~GeV

\end{itemize}

\noindent Here $m_{3trk}$ is the invariant mass of the three tracks in the $Z$ recoil closest to the nominal $a_1(1260)$ mass. See Figure \ref{fig:plots} (left) for the recoil $h_{1,2}$ mass distribution after full analysis selection. 

In order to study the separate mass cases, we apply one of the following recoil mass window requirements: 
\begin{itemize}

\item Case I: require $123 < m_{recoil} < 160$~GeV;

\item Case II: or require $80 < m_{recoil} < 123$~GeV; 

\item Case III: or require none.

\end{itemize}
The event yields for the $h_2$, $h_1$ and $h_2+h_1$ searches are summarized in Table~\ref{tab:sensitivities}. Though the numbers in Table~\ref{tab:sensitivities} are based on the benchmark in Table~\ref{tab:params}, they can be easily rescaled to other scenarios  with different decay branching ratios, as long as the relevant  kinematics is the same.

\begin{table}[tbp]
\begin{center}
\begin{tabular}{||c|c|c|c||} \hline
 & Case I & Case II & Case III \\ \hline
$S$ & 121 & 182 & 302 \\ \hline
$B$ & 0.4 & 1.3 & 1.7 \\ \hline
\end{tabular}
\caption{Signal (S) and background (B) yields for the $h_2$, $h_1$ and $h_2+h_1$ searches in early running at the ILC, respectively. The yields assume $\sqrt{s}=250$~GeV, $250$fb$^{-1}$ integrated luminosity, and 80\% $e_{L}^{-}$, 30\% $e_{R}^{+}$ beam polarization. }   
\label{tab:sensitivities}
\end{center}
\end{table}
The clean environment of an $e^+ e^-$ collider ensures the power of track multiplicity for background suppression. Crossfeed from other $a_1$ decay channels is negligible. The recoil mass distributions are fit with the Gaussian Peak Exponential Tail (GPET) PDF with $m_{h_{1,2}}$ as free parameters. The fits yield $m_{h_{1}}=90.8 \pm 0.2$~GeV and $m_{h_{2}}=124.7 \pm 0.2$~GeV. The information presented in Table~\ref{tab:sensitivities} can be easily rescaled to other points in parameter space.

Luminosity upgrades at $\sqrt{s}=250$~GeV are expected  to yield an additional 1150~fb$^{-1}$ of integrated luminosity \cite{Adolphsen:1541315prime}. Extrapolated to this dataset, the signal event yields are 557, 837 and 1389 for Cases I,II, III respectively .

\subsection{The $h_{1,2} \rightarrow 2a_1 \rightarrow 2 \mu 2\tau$ Channel}

Here we seek to identify $a_1 \rightarrow \mu^+ \mu^-$ events without requiring the $Z \rightarrow \mu^+ \mu^-$ decay channel, greatly enlarging the signal yield. On the $Z$ side we require no-track or two-track decays $Z \rightarrow \nu \bar{\nu}, e^+ e^-, \mu^+ \mu^+, \tau_{1-pr}, \tau_{1-pr}$ and on the $h_{1,2}$ side require one $a_1 \rightarrow \mu^+ \mu^-$ and one $a_1 \rightarrow \tau^+ \tau^-$ where the taus decays as either 1- or 3-prongs:

\begin{itemize}

\item require at least two muons with $p_T>5$~GeV ($N_{\mu5} \geq 2$)

\item require exactly six or eight tracks with $p_T>0.2$~GeV ($N_{trk}=6,8$)

\item require zero net charge in the tracks ($Q_{trks}=0$)

\item require the muon pair mass closest to the $a_1$ mass within $3\sigma$ of the fitted $a_1$ mass ($\vert m_{a_1}-m_{\mu^+ \mu^-} \vert < 3\sigma$)

\end{itemize}

\noindent See Figure \ref{fig:plots} (right) for the distribution of invariant mass of the muon pair closest to the $a_1$ mass. This distribution is fit with a Gaussian PDF for the signal and a linear PDF for the background. Crossfeed from the $2a_1 \rightarrow 4\tau$ channel is flat and is therefore eliminated in the fit.  The fit yields $m_{a_{1}}=10.329 \pm 0.005$~GeV. The expected SM background is 0.7 events and the expected signal yield is 23 events for Case III. After luminosity upgrades, the expected number of signal events is 106 for Case III.

\section{LHC Study}

As previously indicated, neither ATLAS nor CMS have public search results for the benchmark assumed in this study, though CMS has studied  $h \rightarrow 2 a_1$ for $m_{a_1}<2 m_{\tau}$ \cite{Chatrchyan:1490272}. Previous LHC $h \rightarrow 2 a_1 \rightarrow 4\tau$ studies have claimed varying degrees of sensitivity, but do not address the challenges of triggering, background estimation and suppression, and $\tau$ identification with equal vigor. The study \cite{Lisanti:2009uy} does not address background from diboson production in association with jets. The study \cite{Belyaev:2008gj} included neither detector simulation nor background estimation, while \cite{Englert:2011iz} did not include detector simulation and does not address Drell-Yan background. Finally, the study \cite{Cerdeno:2013cz} does not address the important issue of triggering.

Our own preliminary study on the LHC sensitivity suggests the search will be challenging. We use Pythia8 \cite{Sjostrand:2007gs} for event generation at $\sqrt{s}=8$~TeV and the Delphes fast detector simulation \cite{deFavereau:2013fsa}. 
Since the $b$ quark decay to missing energy and soft, oppositely charged leptons which are highly collimated easily imitates light signal $a_1 \rightarrow \tau^+ \tau^-$ decays, any process in which $b$ quarks participate must be accounted. Simulation of background samples with sufficiently large equivalent LHC luminosity is challenging, and such background estimation must be performed in data control samples. Indeed, our studies using the technique of cut-scaling, in which sample efficiency is taken to be the product of exclusive cut efficiencies, rather than cumulative cut efficiencies indicate that the backgrounds $c\bar{c}$ and $b\bar{b}$ may be ineliminable. 
Our studies suggest that the hadronic $\tau$ identification used at CMS \cite{Chatrchyan:2012zz} will need substantive  modification and validation in order to separately identify highly collimated taus, which have mean pair separation of $\Delta R=0.5$. For most hadronic $\tau$ pairs in signal events, one $\tau$ occupies the isolation annulus of the other and vice versa. 
These issues prevent extrapolation to higher luminosity and energy with any degree of confidence. At higher luminosities trigger thresholds must be raised and the impact of pileup on both $\tau$ triggers and offline $\tau$ reconstruction must be mitigated. Nevertheless, we hope our colleagues at the LHC will pursue this promising channel with the data which has already been recorded.

\section{Conclusion}

Exotic Higgs decays may prove to be the window into new physics. But, there exist challenging cases at the LHC, such as the channels with pure MET or soft jets (with or without MET) in the final state. At the ILC, the clean interaction environment provides powerful separation between signal and background processes. We have performed a study of the exotic Higgs decay $h_{1,2} \rightarrow  a_1 a_1$ in the NMSSM with full simulation of the SiD detector at the ILC. The study assumes $\sqrt{s}=250$~GeV with $\int dt \mathcal{L}=250$fb$^{-1}$ after initial running and $\int dt \mathcal{L}=1150$fb$^{-1}$ after luminosity upgrades. 

After initial running with $\int dt \mathcal{L}=250$fb$^{-1}$, we expect discovery for both $h_{125}=h_2$ (SM-like Higgs) and $h_1$ (non-standard Higgs, $m_{h_1}=91.6$GeV). With full SM background simulation, we expect nearly negligible background and approximately 1691 signal events after luminosity upgrades. We find that the expected precision on the $a_1$ mass in early running is $m_{a_{1}}=10.329 \pm 0.005$~GeV as measured in the $2\mu 2\tau$ channel alone, with significant improvement expected after luminosity upgrades.
 It should also be noted that the results here only include the $Z \rightarrow \mu^+ \mu^-$ tag, and that by including the $Z \rightarrow e^+ e^-$ tag the sensitivity should improve by a factor approximately $\sqrt{2}$ modulo small efficiency corrections. Moreover, by including hadronic $Z$ tags the precision will be substantially extended.

In contrast, we find that at the LHC triggering on signal events will be challenging, especially at higher luminosities. Moreover the required modification of $\tau$ identification for identifying highly collimated hadronic $\tau$ pairs in a high pileup environment has not been studied, and the impact on signal sensitivity has not been established to our knowledge.

\section*{Acknowledgments}

Thanks to Tim Barklow for generating the Whizard events and Norman Graf for SiD detector simulation and event reconstruction. Thanks also to the CSS2013 conveners of the Beyond the Standard Model subgroup (Energy Frontier working group): Yuri Gerstein, Markus Luty, Meenakshi Narain, Liantao Wang and Daniel Whiteson.

\bibliography{paper}{}

\end{document}